\documentclass[conference]{IEEEtran}
\IEEEoverridecommandlockouts
\usepackage[table]{xcolor}
\usepackage{algorithm}
\usepackage{lipsum}  
\usepackage{cite}
\usepackage{dblfloatfix}
\usepackage{float}
\usepackage{comment}
\usepackage{amsmath,amssymb,amsfonts}
\usepackage{mathtools}
\usepackage{algorithm}
\usepackage{steinmetz}
\usepackage[noend]{algpseudocode}
\usepackage{graphicx}
\graphicspath{{./Figures/}}
\usepackage{textcomp}
\usepackage{siunitx}
\usepackage{paralist} 
\usepackage{url}
\usepackage{array,multirow}
\usepackage[table]{xcolor}
\usepackage{xcolor,colortbl}
\def\BibTeX{{\rm B\kern-.05em{\sc i\kern-.025em b}\kern-.08em
T\kern-.1667em\lower.7ex\hbox{E}\kern-.125emX}}

\begin{document}	
\title{Real-Time T\&D Co-Simulation for Testing Grid Impact of High DER Participation}

\author{\IEEEauthorblockN{Victor Paduani, Rahul Kadavil, Hossein Hooshyar}
\IEEEauthorblockA{Advanced Grid Innovation Laboratory for Energy} 
New York Power Authority, White Plains, NY\\victor.daldeganpaduani@nypa.gov
\and
\IEEEauthorblockN{Aboutaleb Haddadi$^{\dagger}$, AHM Jakaria$^{\ddagger}$, Aminul Huque$^{\ddagger}$}
\IEEEauthorblockA{$^{\dagger}$Transmission Ops. and Planning, $^{\ddagger}$DER Integration}
Electric Power Research Institute, Knoxville, TN\\ahaddadi@epri.com

\thanks{This  research  is  supported  by  the  U.S.  Department  of  Energy’s  Office  of Energy  Efficiency  and  Renewable  Energy  (EERE)  under  the  Solar  Energy Technologies Office Award Number DE-EE0009021.}
}

\maketitle

\begin{abstract}

\textbf{This paper presents the development of a real-time T\&D co-simulation testbed for simulating large grids under high DER penetration. By integrating bulk power system, distribution feeders, and distributed energy resources (DER) models into one simulation environment, the testbed enables the performance analysis and validation of DER management systems (DERMS) algorithms. This work proposes a co-simulation timestep sequence for the cross-platform data exchange and time synchronization, with a communication framework based on MQTT communication protocol. The proposed strategy is tested with a 5,000 buses model of part of the North American bulk power system (BPS) and a 9,500 nodes distribution feeder model obtained from a local utility. Simulations are carried out to demonstrate the capability of the proposed framework to propagate events between the transmission and distribution models. Results are used to quantify how the co-simulation timestep size can affect the propagation of dynamics between the models.}

\end{abstract}


\section{Introduction}

In 2020, the Federal Energy Regulation Commission (FERC) approved Order 2222, envisioning the development of the future grid, in which distributed energy resources (DERs) can participate in wholesale energy markets through aggregators. This new rule, in effect since February 2022, allows DERs to be organized and controlled to provide grid services, improving the grid flexibility and resilience \cite{cano2020ferc}. Moreover, to increase the interoperability of DERs, the IEEE standard 1547-2018 \cite{ieee1547} requires every DER to provide basic grid-support functionalities (GSFs) such as frequency-watt and volt-var droop response, disturbance ride through, and so forth. Consequently, there has been a growing interest from both transmission and distribution operators to utilize utility-scale and behind-the-meter (BTM) DERs to provide grid services \cite{ardani2018coordinating}, \cite{aminul2021grid}. 

\begin{figure}[htb]
    \centering
    \includegraphics[width=0.5\textwidth]{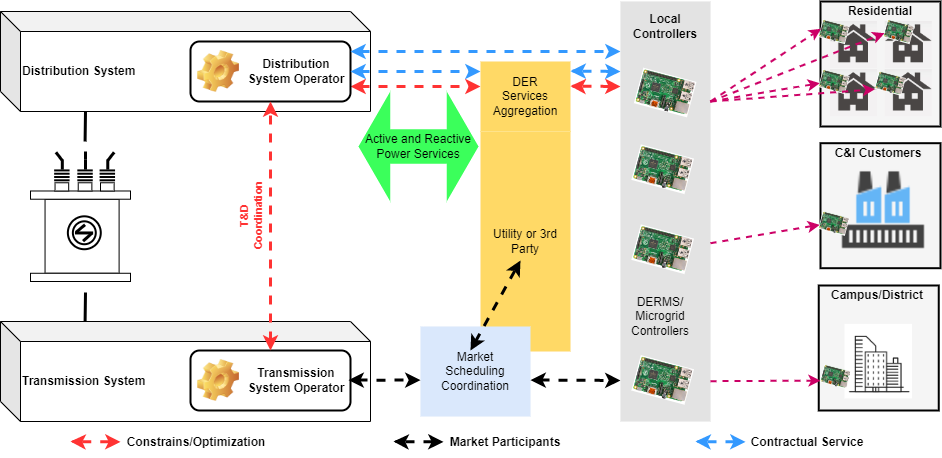}
    \caption{DERMS framework proposed in \cite{aminul2021distributed} as part of the `\textit{Enable BTM DER-provided Grid Services that Maximize Customer Grid Benefits}' (ENGAGE) project.}
    \label{fig:engage_intro}
\end{figure}

In \cite{aminul2021distributed}, Aminul \textit{et al}. introduce a DER management system (DERMS) for enabling BTM solar co-located with other DERs to provide bulk power system (BPS) and distribution system services to improve system reliability and economic efficiency. Figure \ref{fig:engage_intro} displays an overview of the proposed strategy, in which transmission and distribution operators coordinate to request grid services. As shown in the figure, utility-scale DERs and/or multiple individual BTM DERs are managed by local controllers, which are then combined together by aggregators. Thus, the aggregator can be used to provide services to grid operators and/or wholesale market operators.  

Since BPS-level grid services provision by BTM DERs impact distribution, and vice-versa, T\&D co-simulations are an attractive approach for validating their performance both in steady-state and/or dynamic studies. Cross-platform co-simulation techniques are an active research topic that has significantly advanced in recent years \cite{palmintier2017design, hansen2016enabling}. In \cite{jain2021integrated}, Jain \textit{et al}. conduct an extensive review of T\&D co-simulation approaches found in the literature, observing that there is still no consensus on what approach should be adopted across the industry. Moreover, co-simulation in real-time presents more challenging aspects due to the needs for real-time synchronization and operating the system without overruns, which occur when the model takes longer than its real-time timestep to calculate the next iteration solution \cite{paduani2021novel}. 
To address the data exchange and time synchronization challenges of real-time co-simulations, in this work, we present the development of a real-time T\&D co-simulation testbed that combines transient stability (TS) models of large BPS, quasi-static time-series (QSTS) models of multiple distribution feeders, and DER models developed to provide GSFs in accordance to grid standards \cite{ieee1547}, \cite{calrule21}. The coupling is performed via MQTT communication protocol, particularly useful for internet of things (IoT) applications \cite{atmoko2017iot}. The contributions of this work to the literature are summarized as follows.

\begin{itemize}
    \item Presents the development of a real-time T\&D co-simulation testbed that enables the performance analysis and validation of DERMS algorithms.
    \item Analyzes the propagation of dynamics between the transmission and distribution models with the proposed co-simulation communication architecture.
\end{itemize}

\section{Methodology}

\subsection{Co-simulation Coupling and Architecture}

\begin{figure}[htb]
    \centering
    \includegraphics[width=0.5\textwidth]{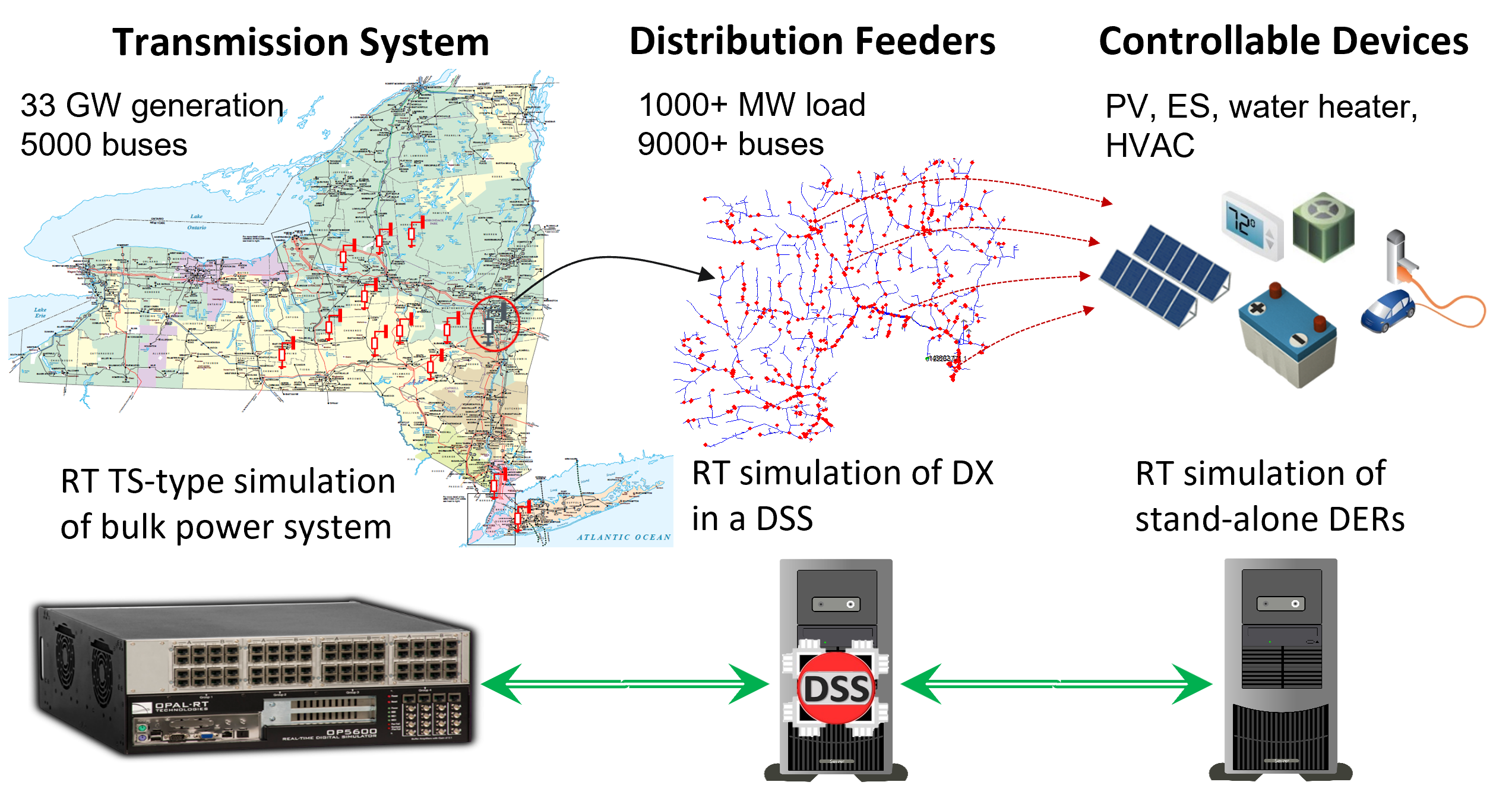}
    \caption{Proposed T\&D co-simulation framework.}
    \label{fig:cosim_struct}
\end{figure}

The co-simulation structure is presented in Fig. \ref{fig:cosim_struct}. The system consists of (i) a section of the Eastern Interconnection Transmission System (EITS), containing 33 GW of generation and over 5,000 buses ranging from 69 to 765 kV, (ii) realistic distribution feeder models obtained from local utilities with loads in the range of thousands of MW and over 9000 nodes each, and (iii) controllable devices including photovoltaic (PV), energy storage (ES), water heaters, and HVAC. The TS model of the EITS is simulated in a TS-type real-time simulator (RTS), the distribution feeder models (originally built in Milsoft) are simulated in the distribution system simulator (DSS), and the DERs are implemented by a stand-alone DER simulator software responsible for calculating their corresponding PQ injections based on grid information at their point of interconnection (POI) and the GSF enabled for each device. 

\subsection{Data Exchange and Time Synchronization}
The co-simulation communication architecture, based on MQTT communication protocol, is encapsulated by functions written in Python programming language, which enable cross-platform data exchange within a single system and can be scalable across multiple computational architecture. Note MQTT is a Client-Server publish/subscribe messaging transport protocol ideal for IoT applications due to its light weight and easy implementation \cite{standard2014mqtt}. By providing a simplified abstraction mechanism to create lightweight and agnostic communication packets for asynchronous data exchange, the MQTT protocol allows (i) reliable data delivery, (ii) data broadcasting to a group of devices with relative ease, and (iii) proven scalability with millions of devices. 

In the proposed architecture, the RTS and DSS are simulated as standalone MQTT client applications that can publish and subscribe measurement data. The RTS and DSS clients are tied to the MQTT server (also referred to as the broker) closing the data-exchange communication loop. The broker serves as the central hub for handling all MQTT messages from its connected clients, receiving and broadcasting the messages to the appropriate clients over a network.

The coupling between the different simulation tools, displayed in Fig. 2, is implemented as follows. First, the RTS (on the left) operates as the co-simulation master controller, dictating the global timestamp, and initializing the operation of each co-simulation timestep. The timestamp reference from the RTS is synchronized with the MQTT Server using communication adapter functions built in Python, which are ran in an external computer. Next, measurement information consisting of voltage and frequency at a POI in the EITS model are sent from the RTS to the DSS, whereas active and reactive power load consumption from the feeder(s) head (simulated in the DSS) are sent back as spot loads to the EITS model (simulated in RTS). Then, at the beginning of each co-simulation timestep, the RTS client updates the voltage and frequency measurements as well as simulation time from each DER’s defined POI and publishes the data to the MQTT broker. Once the data is received by the broker, it notifies the DSS client as well as the stand-alone DER simulator client (subscribers). 

During real-time operation, the DSS and stand-alone DER simulator clients will work in tandem to perform: (i) an unbalanced power flow (PF) to find the new optimal operating points of the distribution system, and (ii) solve for the new DER power injections based on the power flow solution and their corresponding GSFs settings. The results from the DSS client are then published back to the RTS client via MQTT broker and the co-simulation loop is completed. It is worth mentioning that the implementation details for the real-time data exchange and time synchronization in development for this testbed will be presented in detail in future works.

\section{Simulation Results}

In this section, we present the following simulation results: (A) dynamics propagation delay that highlights how events propagate between the BPS and the distribution feeders; (B) frequency spectrum propagation limits, in which we demonstrate how the co-simulation timestep can limit the maximum frequency of events propagating between the platforms; (C) co-simulation loop delay, and (D) DER interconnection dynamics. 

\begin{figure}[htb]
    \centering
    \includegraphics[width=0.5\textwidth]{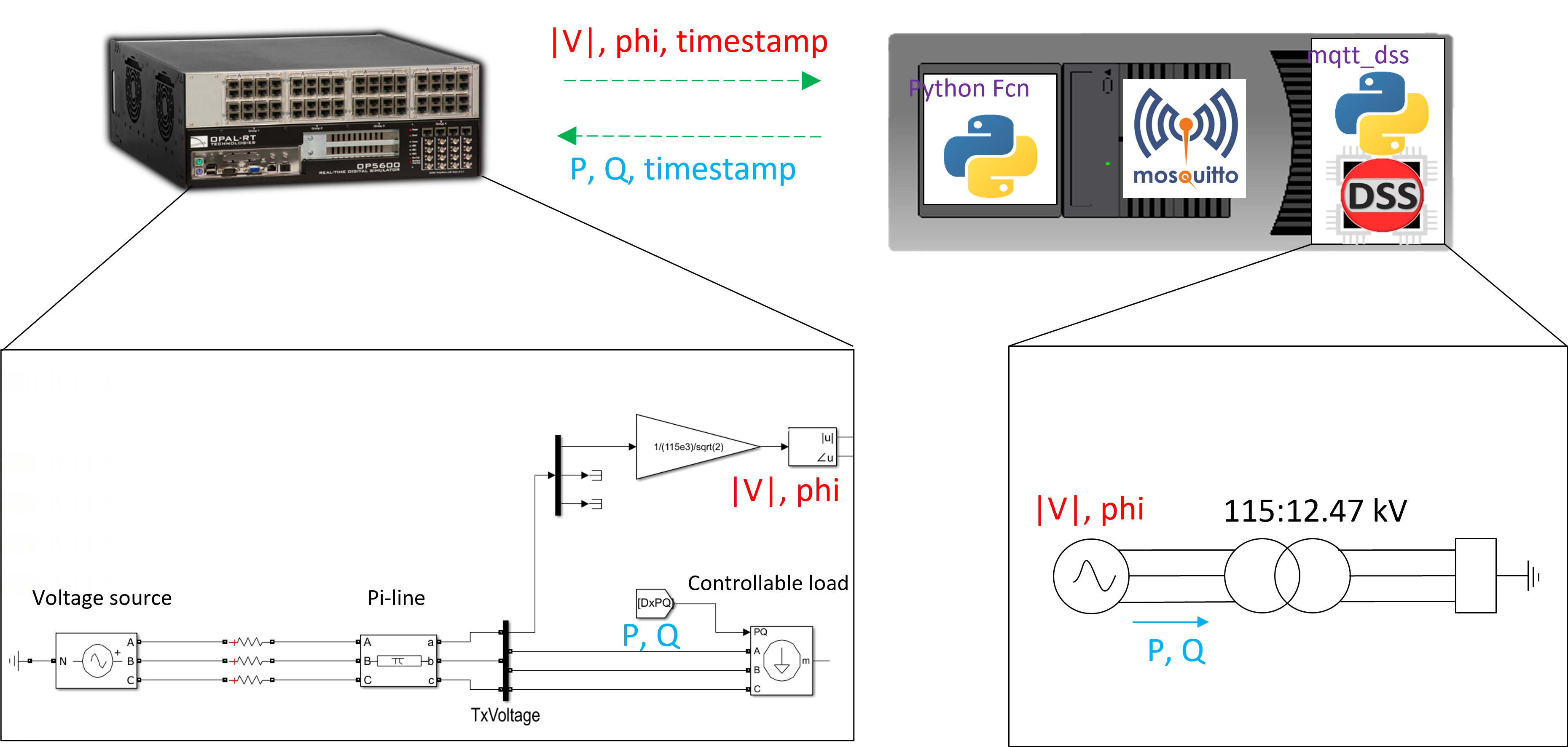}
    \caption{Prototype built for tests (A) and (B).}
    \label{fig:prototype}
\end{figure}

A small prototype of the real-time T\&D co-simulation testbed is built for obtaining results (A) and (B). As displayed in Fig. \ref{fig:prototype}, the prototype consists of a voltage source representing the transmission system connected via a pi-line to a controllable load representing the distribution system. The simplification is made to focus on the data exchange and time synchronization features. This is because those results are related to the propagation limits associated with the co-simulation timestep, and are independent from the size of the transmission and distribution systems. Three versions of the prototype model are built for comparing the results: (i) with the proposed co-simulation platform, (ii) entirely in the RT simulator, and (iii) entirely in the DSS.

For simulation results (C) and (D), the 5,000 nodes TS-type RTS model of the EITS is co-simulated with a 9,500 nodes distribution feeder DSS model (Fig. \ref{fig:cosim_struct}). The BPs model utilizes 10 cores to run in real-time without overruns with a 5 ms timestep. More details on the parallelized simulation are discussed in \cite{eph2022userguide}. The distribution feeders are simulated in DSS in an external computer.

\subsection{Dynamics Propagation Delay}

\begin{figure}[htb]
    \centering
    \includegraphics[width=0.5\textwidth]{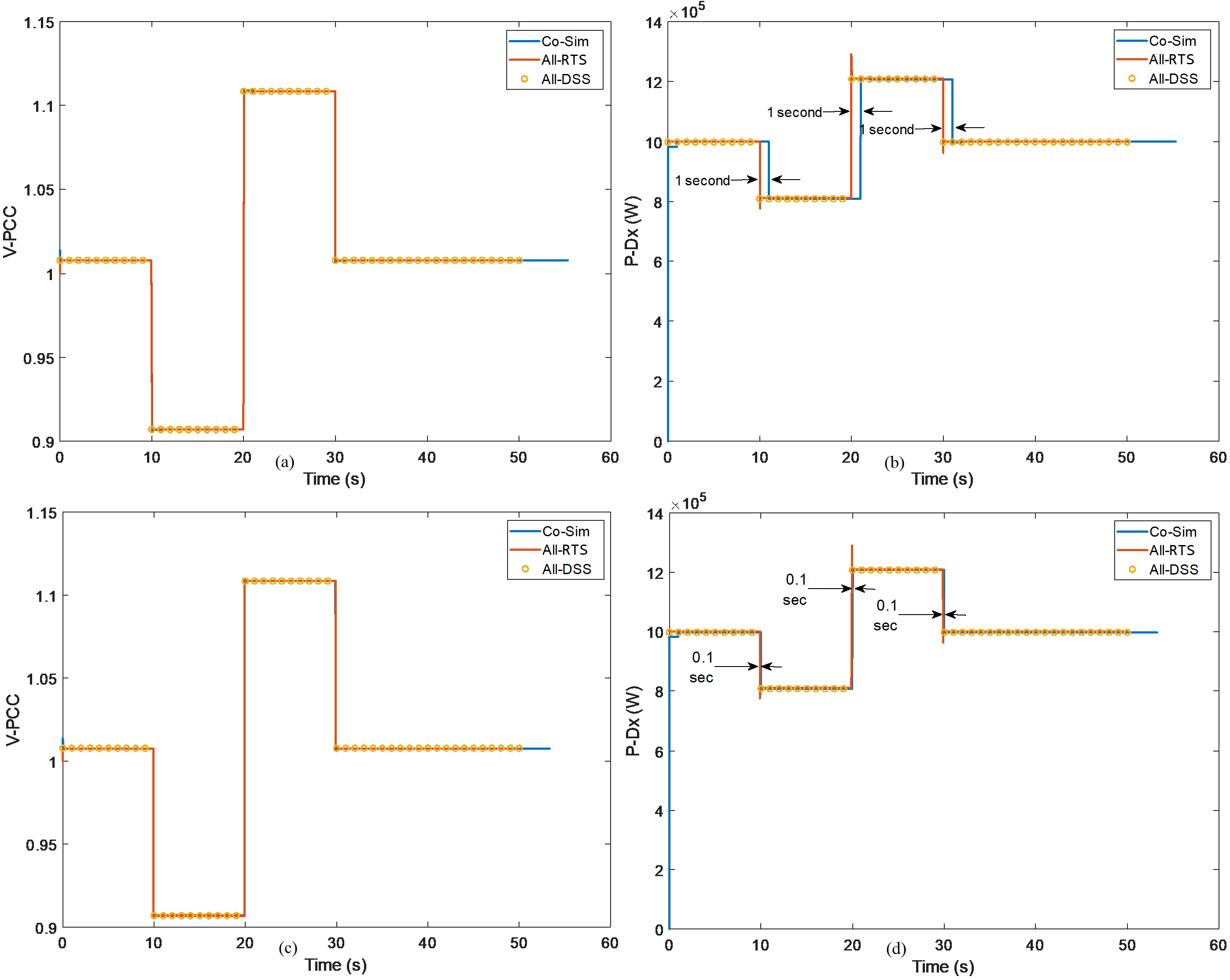}
    \caption{Impact of co-simulation loop delay on the propagation of dynamics between the TS-type RTS and the DSS: (a) Voltage in scenario 1; (b) Load active power in scenario 1; (c) Voltage in scenario 2; (d) Load active power in scenario 2.}
    \label{fig:scenario1and2}
\end{figure}

This test is divided into two scenarios. In the first, the voltage reference of the transmission bus to which the distribution feeder is connected to is configured to perform a step change at the same moment when the co-simulation timestep occurs. Then, we analyze the voltage at the transmission-distribution point of common coupling (V-PCC) and the active power consumed by the distribution system (P-Dx). Figure \ref{fig:scenario1and2} compares the results for the distribution feeder load consumption between the prototype models.

As shown in Fig. \ref{fig:scenario1and2}b, the P-Dx obtained from the co-simulation setup displays a 1-second delay when compared to its counterparts built entirely in the RTS or entirely in the DSS. This happens because the voltage reference step occurred when the transmission data had already been transmitted via UDP to the distribution simulation in DSS, causing the distribution system to only receive the updated PCC voltage in the next co-simulation timestep. 
In the second scenario, the PCC voltage reference is configured to perform the step change 0.1 second before the next co-simulation timestep. Figure \ref{fig:scenario1and2}d shows that, in this case, the propagation delay between the transmission and distribution was reduced to 0.1 second. This demonstrates the maximum delay that may occur for the propagation of dynamics between the RTS and DSS is proportional to the co-simulation closed-loop delay, which should be designed as small as possible without causing overruns.

\subsection{Frequency Spectrum Propagation Limits}

\begin{figure}[htb]
    \centering
    \includegraphics[width=0.5\textwidth]{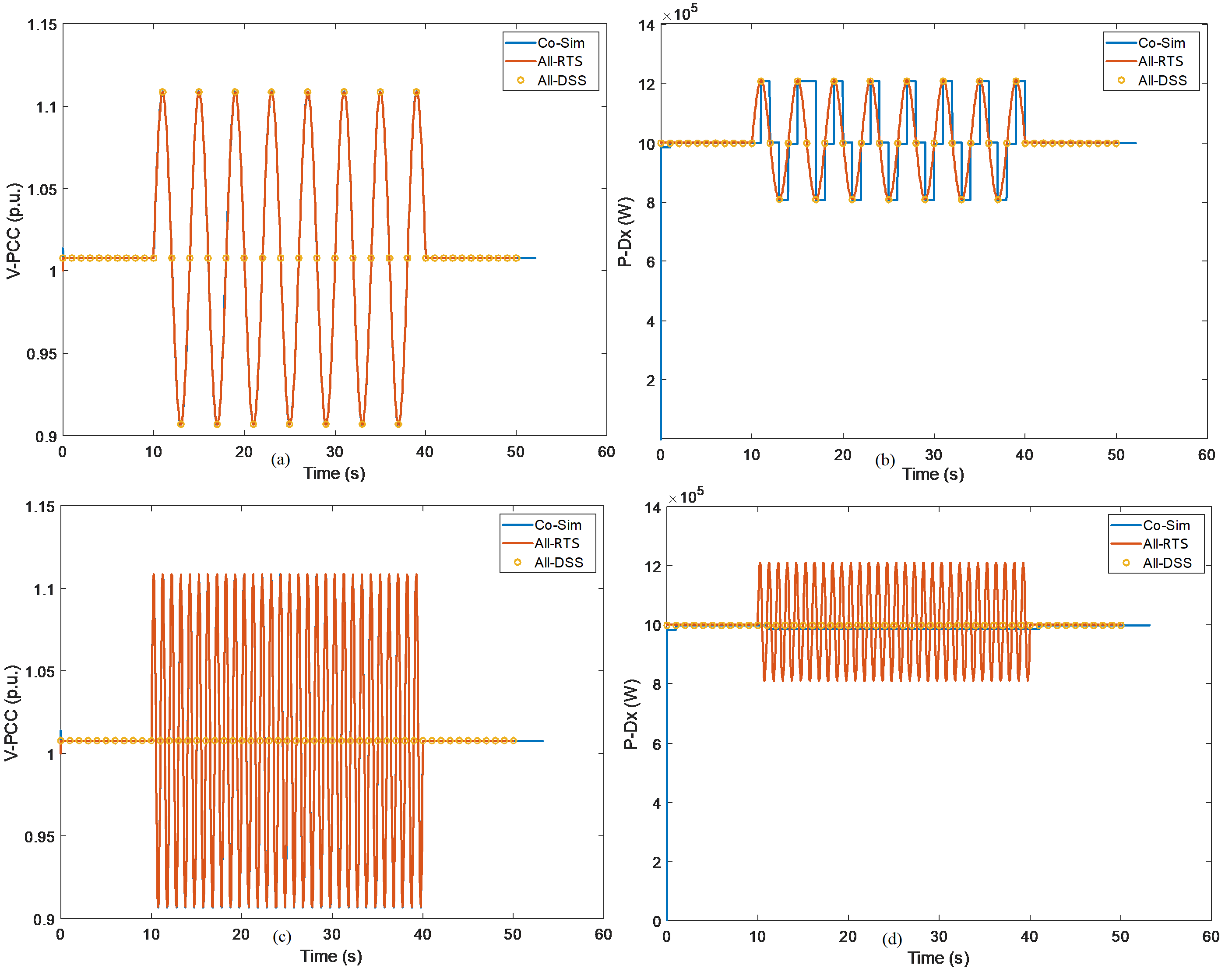}
    \caption{Frequency spectrum propagation limits between the TS-type RTS and the DSS: (a) Voltage in scenario 3; (b) Load active power in scenario 3; (c) Voltage in scenario 4; (d) Load active power in scenario 4.}
    \label{fig:scenario3c}
\end{figure}

For scenario 3, we analyze the frequency spectrum propagation limits of the dynamics between the transmission and distribution simulations. In this case, the voltage magnitude of the transmission voltage source is modulated by a sinusoidal waveform of 0.1 p.u. magnitude and 0.25 Hz frequency. Figures \ref{fig:scenario3c}a and \ref{fig:scenario3c}b show the voltage and load consumption, respectively. From the results it can be seen how the voltage dynamics are able to propagate to the distribution system, directly affecting the feeder load consumption.

Next, for scenario 4, the modulated sinusoidal waveform frequency is increased from 0.25 to 1 Hz. Results are displayed in Figs. \ref{fig:scenario3c}c and \ref{fig:scenario3c}d. In this case, the voltage at the transmission bus properly follows the reference; however, the load consumed by the distribution system (Fig. \ref{fig:scenario3c}) is not able to track the waveform from the co-simulation, causing its results to divergence from the results obtained when simulating the entire scenario within RTS. That occurred because the modulation signal applied to the voltage reference had a frequency value above the maximum frequency spectrum that could propagate between the domains.

It is found that the limit frequency for the propagation of events between the transmission and distribution models is lesser than half the frequency of the co-simulation timestep, e.g., ($f < 0.5/T_{s}$). This highlights the need for reducing the co-simulation timestep as possible.

\subsection{Closed-loop Delay Distribution}

\begin{figure}[htb]
    \centering
    \includegraphics[width=0.4\textwidth]{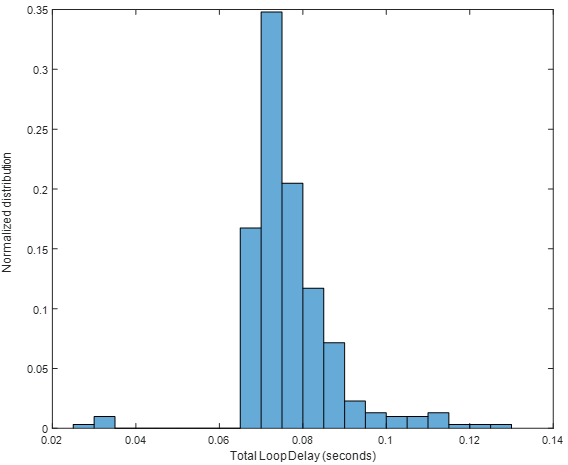}
    \caption{Statistics of closed-loop delay collected over 10 minutes of operation.}
    \label{fig:closedloopdelay}
\end{figure}

It is important to design the co-simulation closed-loop timestep to be larger than the time it takes for the DSS power flow to converge after N iterations. More specifically, the co-simulation timestep must be always larger than the closed-loop delay between the RTS and the DSS. Figure \ref{fig:closedloopdelay} demonstrates the distribution of the time required to complete steps (1) to (5) from the co-simulation timestep sequence for 10 minutes of operation. Note the test is carried out with the EITS integrated with a 9,500 nodes distribution system with 1000 DERs. 

Results demonstrate that the co-simulation closed-loop delay requires an average of 80 ms and a maximum value considerably below the co-simulation timestep of 1 second, hence achieving the co-simulation platform requirements.

\subsection{DER Interconnection Dynamics}

\begin{figure}[htb]
    \centering
    \includegraphics[width=0.5\textwidth]{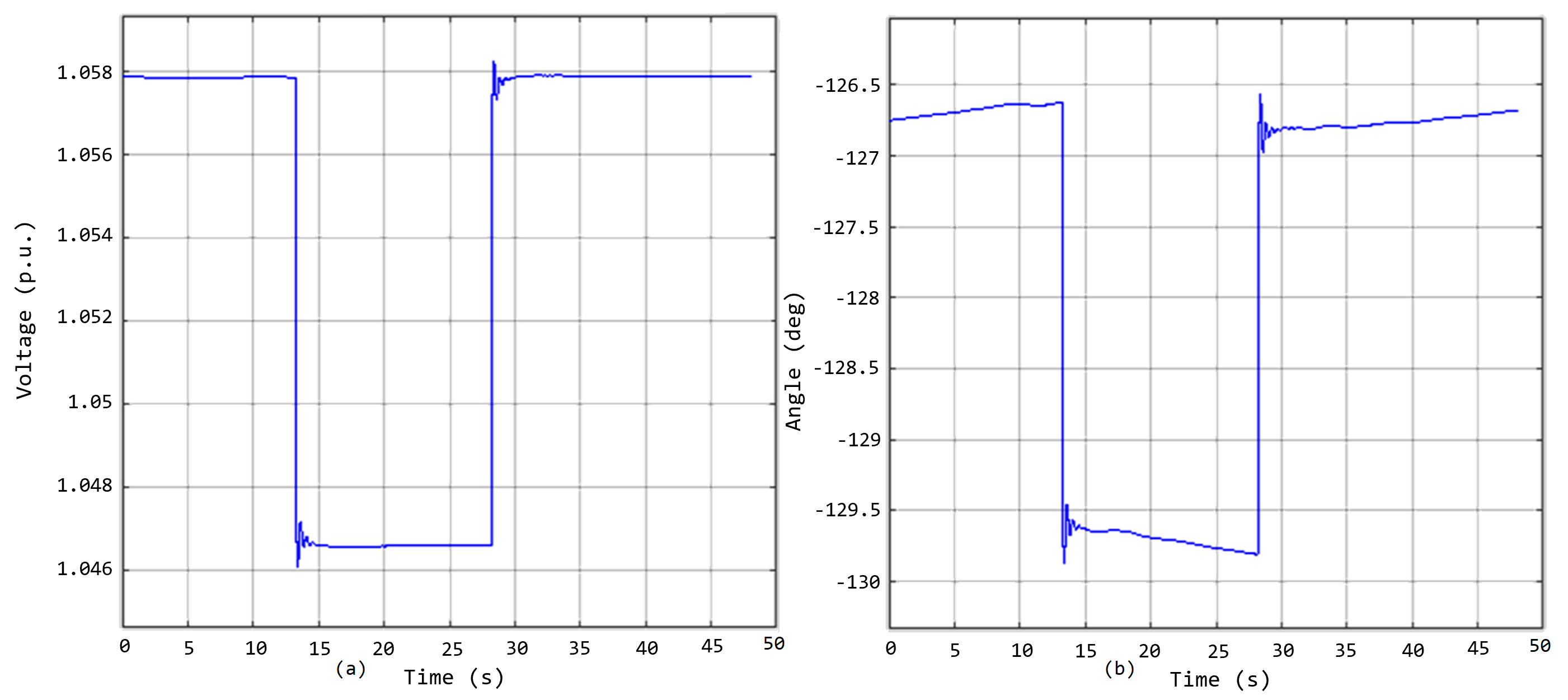}
    \caption{Transients during DER connection/disconnection events: (a) RMS positive sequence voltage amplitude, (b) phase angle}
    \label{fig:vdynamics}
\end{figure}

Next, we present some preliminary results for the interconnection of 1000 BTM DERs to the 9,500 nodes distribution feeder. Figure \ref{fig:vdynamics} shows voltage RMS transients at the transmission bus connected to the feeder head (simulated in a TS-type RTS) caused by the connection/disconnection of DERs to/from the distribution feeder (simulated in the DSS). First, around t = 13 seconds, the DERs are disconnected, and we can observe a voltage drop of approximately 0.013 p.u., displayed in Fig. \ref{fig:vdynamics}. Then, around second 27, the DERs are reconnected to the system, boosting the feeder head voltage back to its previous state. Note the dynamics are fast in this scenario because (i) they are purely electrical (no motor loads were included), and (ii) the DERs update their output in steps (no ramp rate effect included). The results demonstrate how events occurring at the distribution level can propagate via the co-simulation communication architecture (Fig. \ref{fig:cosim_struct}) to impact the voltage of a bus at the transmission level. Tests including ramp rate and motor load will be executed in future work. 

\section{Conclusion}

In this paper, we present the development of a real-time T\&D co-simulation framework for simulating large grids under high penetration of DERs. The system utilizes MQTT protocol for the co-simulation data exchange and time-synchronization. Results demonstrate that depending on the moment an event occurs, there can be a delay up to the co-simulation timestep for the dynamics to propagate from one simulation software to another. Furthermore, it was demonstrated that the maximum frequency spectrum that can be analyzed by the platform is determined by half the size of the co-simulation timestep. We also demonstrate the dynamics propagation to the EITS model from interconnecting 1000 DERs to a 9,500 distribution feeder. In future work, the testbed will be used to analyze the impacts of substituting large synchronous generators from the EITS by distributed BTM DERs capable of providing grid services.


\bibliographystyle{IEEEtran}
\bibliography{mybibtex}

\begin{thebibliography}{10}
\providecommand{\url}[1]{#1}
\csname url@samestyle\endcsname
\providecommand{\newblock}{\relax}
\providecommand{\bibinfo}[2]{#2}
\providecommand{\BIBentrySTDinterwordspacing}{\spaceskip=0pt\relax}
\providecommand{\BIBentryALTinterwordstretchfactor}{4}
\providecommand{\BIBentryALTinterwordspacing}{\spaceskip=\fontdimen2\font plus
\BIBentryALTinterwordstretchfactor\fontdimen3\font minus
  \fontdimen4\font\relax}
\providecommand{\BIBforeignlanguage}[2]{{%
\expandafter\ifx\csname l@#1\endcsname\relax
\typeout{** WARNING: IEEEtran.bst: No hyphenation pattern has been}%
\typeout{** loaded for the language `#1'. Using the pattern for}%
\typeout{** the default language instead.}%
\else
\language=\csname l@#1\endcsname
\fi
#2}}
\providecommand{\BIBdecl}{\relax}
\BIBdecl

\bibitem{cano2020ferc}
C.~Cano, ``{FERC Order No. 2222: A New Day for Distributed Energy Resources},''
  2020.

\bibitem{ieee1547}
``{IEEE Standard 1547-2018},'' \emph{Standard for interconnection and
  interoperability of distributed energy resources with associated electric
  power systems interfaces}, 2018.

\bibitem{ardani2018coordinating}
K.~B. Ardani, E.~J. O'Shaughnessy, and P.~D. Schwabe, ``Coordinating
  distributed energy resources for grid services: A case study of pacific gas
  and electric,'' National Renewable Energy Lab.(NREL), Golden, CO (United
  States), Tech. Rep., 2018.

\bibitem{aminul2021grid}
N.~Singhal, M.~Heidarifar, T.~Hubert, E.~Ela, A.~Huque, T.~Abate, and P.~Shoop,
  ``Grid services in the distribution and bulk power systems,'' Electric Power
  Research Institute.(EPRI), Palo Alto, CA (United States), Tech. Rep., 2021.

\bibitem{aminul2021distributed}
A.~Garg, T.~Hubert, A.~Huque, and A.~Renjit, ``Distributed energy resource
  magagement system ({DERMS}) control architecture for grid services in the
  distribution and bulk power systems.'' Electric Power Research
  Institute.(EPRI), Palo Alto, CA (United States), Tech. Rep., 2021.

\bibitem{palmintier2017design}
B.~Palmintier, D.~Krishnamurthy, P.~Top, S.~Smith, J.~Daily, and J.~Fuller,
  ``Design of the {HELICS} high-performance
  transmission-distribution-communication-market co-simulation framework,'' in
  \emph{2017 Workshop on Modeling and Simulation of Cyber-Physical Energy
  Systems (MSCPES)}.\hskip 1em plus 0.5em minus 0.4em\relax IEEE, 2017, pp.
  1--6.

\bibitem{hansen2016enabling}
T.~M. Hansen, R.~Kadavil, B.~Palmintier, S.~Suryanarayanan, A.~A. Maciejewski,
  H.~J. Siegel, E.~K. Chong, and E.~Hale, ``Enabling smart grid cosimulation
  studies: Rapid design and development of the technologies and controls,''
  \emph{IEEE Electrification Magazine}, vol.~4, no.~1, pp. 25--32, 2016.

\bibitem{jain2021integrated}
H.~Jain, B.~A. Bhatti, T.~Wu, B.~Mather, and R.~Broadwater, ``Integrated
  transmission-and-distribution system modeling of power systems:
  State-of-the-art and future research directions,'' \emph{Energies}, vol.~14,
  no.~1, p.~12, 2021.

\bibitem{paduani2021novel}
V.~Paduani, B.~Xu, D.~Lubkeman, and N.~Lu, ``Novel real-time {EMT-TS} modeling
  architecture for feeder blackstart simulations,'' in \emph{2022 IEEE Power \&
  Energy Society General Meeting (PESGM)}.\hskip 1em plus 0.5em minus
  0.4em\relax IEEE, 2022.

\bibitem{calrule21}
{California Public Utilities Commission}, ``{Electric Rule No. 21 Generating
  facility interconnections},'' 2016.

\bibitem{atmoko2017iot}
R.~Atmoko, R.~Riantini, and M.~Hasin, ``{IoT} real time data acquisition using
  {MQTT} protocol,'' in \emph{Journal of Physics: Conference Series}, vol. 853,
  no.~1.\hskip 1em plus 0.5em minus 0.4em\relax IOP Publishing, 2017, p.
  012003.

\bibitem{standard2014mqtt}
\BIBentryALTinterwordspacing
``{MQTT} version 3.1.1.'' [Online]. Available:
  \url{http://docs.oasis-open.org/mqtt/mqtt/v3}
\BIBentrySTDinterwordspacing

\bibitem{eph2022userguide}
\BIBentryALTinterwordspacing
``{ePHASORSIM User guide},'' 2022. [Online]. Available:
  \url{https://wiki.opal-rt.com/display/EUD/Parallel+Simulation}
\BIBentrySTDinterwordspacing

\end{thebibliography}

\end{document}